\documentclass[sigconf, nonacm, screen, pdfa]{acmart}

\newcommand\vldbdoi{10.14778/3836663.3836726}
\newcommand\vldbpages{3799 - 3806}
\newcommand\vldbvolume{19}
\newcommand\vldbissue{11}
\newcommand\vldbyear{2026}
\newcommand\vldbauthors{\authors}
\newcommand\vldbtitle{\shorttitle} 
\newcommand\vldbavailabilityurl{https://github.com/deem-data/stratum}
\newcommand\vldbpagestyle{empty} 

\usepackage{booktabs} 
\usepackage{amsmath}
\usepackage{graphicx}
\usepackage{wrapfig}
\usepackage{subfigure}
\usepackage{subcaption}
\usepackage[np,autolanguage]{numprint}
\usepackage{units}
\usepackage{fontawesome}
\usepackage{balance}
\usepackage{algorithm}
\usepackage[noend]{algpseudocode}
\usepackage{listings}
\usepackage{color}
\usepackage{appendix}
\usepackage{enumitem}
\usepackage{makecell}
\usepackage{cleveref}

\definecolor{dkgreen}{rgb}{0,0.6,0}
\definecolor{gray}{rgb}{0.5,0.5,0.5}
\definecolor{mauve}{rgb}{0.58,0,0.82}
\definecolor{dkred}{rgb}{1.0,0.01,0.24}

\newcommand{\gb}{\unit{\,GB}}

\newcommand{\header}[1]{\vspace{0.7mm}\textbf{#1}}

\newcommand{\workload}{agentic pipeline search}
\newcommand{\Workload}{Agentic Pipeline Search}

\setlist[itemize]{topsep=2pt,itemsep=0pt,parsep=0pt,partopsep=0pt} 

\usepackage[framemethod=default]{mdframed}
\newmdenv[
  backgroundcolor=blue!10,
  linewidth=0pt,
  leftmargin=0pt,
  rightmargin=0pt,
  innerleftmargin=3pt,
  innerrightmargin=3pt,
  innertopmargin=3pt,
  innerbottommargin=3pt,
  skipabove=0.4\baselineskip,
  skipbelow=0\baselineskip,
  splittopskip=\baselineskip,
  splitbottomskip=\baselineskip,
]{reviewbox}

\begin{document}

\title{stratum: A System Infrastructure for Massive~Agent-Centric~ML~Workloads}

\author{Arnab Phani}
\affiliation{\institution{BIFOLD \& TU Berlin}}
\email{arnab.phani@tu-berlin.de}

\author{Elias Strauss}
\affiliation{\institution{BIFOLD \& TU Berlin}}
\email{elias.strauss@tu-berlin.de}

\author{Sebastian Schelter}
\affiliation{\institution{BIFOLD \& TU Berlin}}
\email{schelter@tu-berlin.de}

\renewcommand{\shortauthors}{Arnab Phani et al.}

\begin{abstract}
Recent advances in large language models (LLMs) transform how machine learning (ML) pipelines are developed and evaluated. LLMs enable a new type of workload, \emph{\workload{}}, in which autonomous or semi-autonomous agents generate, validate, and optimize complete data science pipelines. These agents predominantly operate over popular Python ML libraries and exhibit highly exploratory behavior. This results in thousands of executions for data profiling, pipeline generation, and iterative refinement of pipeline stages and hyperparameters.
However, the existing Python-based ML ecosystem is built around libraries such as Pandas and scikit-learn, which are designed for human-centric, interactive, sequential workflows and remain constrained by Python’s interpretive execution model, library-level isolation, and limited runtime support for executing large numbers of pipelines. Meanwhile, many high-performance ML systems proposed by the systems community either target narrow workload classes or require specialized programming models, which limits their integration with the Python ML ecosystem and makes them largely ill-suited for adoption by LLM-based agents. This growing mismatch exposes a fundamental systems challenge in supporting \workload{} at scale.

We therefore propose \emph{stratum}, a unified system infrastructure that decouples pipeline execution from planning and reasoning during \workload{}. Stratum integrates seamlessly with existing Python libraries, compiles batches of agent- or human-generated pipelines into optimized execution graphs, and efficiently executes them across heterogeneous backends, including a novel Rust-based runtime.
We present stratum's architectural vision along with an early prototype, discuss key design decisions, and outline open challenges and research directions. Finally, preliminary experiments show that stratum can significantly speed up large-scale \workload{} up to~16.6x.
\end{abstract}

\maketitle

\pagestyle{\vldbpagestyle}
\begingroup\small\noindent\raggedright\textbf{PVLDB Reference Format:}\\
\vldbauthors. \vldbtitle. PVLDB, \vldbvolume(\vldbissue): \vldbpages, \vldbyear.\\
\href{https://doi.org/\vldbdoi}{doi:\vldbdoi}
\endgroup
\begingroup
\renewcommand\thefootnote{}\footnote{\noindent
This work is licensed under the Creative Commons BY-NC-ND 4.0 International License. Visit \url{https://creativecommons.org/licenses/by-nc-nd/4.0/} to view a copy of this license. For any use beyond those covered by this license, obtain permission by emailing \href{mailto:info@vldb.org}{info@vldb.org}. Copyright is held by the owner/author(s). Publication rights licensed to the VLDB Endowment. \\
\raggedright Proceedings of the VLDB Endowment, Vol. \vldbvolume, No. \vldbissue\ %
ISSN 2150-8097. \\
\href{https://doi.org/\vldbdoi}{doi:\vldbdoi} \\
}\addtocounter{footnote}{-1}\endgroup

\ifdefempty{\vldbavailabilityurl}{}{
\vspace{.3cm}
\begingroup\small\noindent\raggedright\textbf{PVLDB Artifact Availability:}\\
The source code, data, and/or other artifacts have been made available at \url{\vldbavailabilityurl}.
\endgroup
}


\section{Introduction} \label{sec:intro}
The rapid adoption of ML has a profound impact on many domains. Despite this progress, developing ML pipelines remains a labor-intensive process for data scientists, requiring extensive domain knowledge, data engineering effort, iterative experimentation, and exploratory analysis \cite{PolyzotisRWZ18, XinMPP21, SculleyHGDPECYC15}. Recent research has therefore focused on LLM-backed machine learning engineering (MLE) agents \cite{Nam2025, GuoD0C0024, abs-2509-21825, abs-2502-13138, fang2025mlzero, abs-2506-13131}, leveraging their code generation and reasoning capabilities. These agents frame ML tasks as code optimization problems, and explore a large space of candidate solutions by generating and executing a vast number of Python pipelines, guided by their predictive performance on held-out data.

\header{Agentic Workloads:} Large enterprises are increasingly adopting MLE agents for data science and ML application development \cite{Anaconda2024, Cloudera2025}, making \emph{agentic AI} a prominent research direction in the ML community \cite{icmlinsight, neuripsinsight}. Practitioners now employ LLMs across a spectrum of autonomy \cite{abs-2510-23587}---from fully autonomous agents that generate and validate complete pipelines \cite{abs-2502-13138, abs-2509-06503, abs-2509-21825, fang2025mlzero, abs-2506-13131}, to semi-automated frameworks that offload specific pipeline stages \cite{Hollmann0H23, FanFTCLD25, ovcharenko2026sempipes}, to AI-assisted programming, where engineers manually assemble LLM-suggested components using declarative APIs from libraries such as skrub~\cite{skrub} or scikit-learn~\cite{PedregosaVGMTGBPWDVPCBPD11}.
Fully autonomous agents typically translate natural language tasks into executable code through iterative cycles of planning, implementation, and evaluation. These cycles begin with data profiling \cite{abs-2509-21825, fang2025mlzero} and proceed to targeted refinement of ML pipeline stages \cite{Nam2025, abs-2506-13131}, seamlessly combining widely used libraries with specialized solutions for both tabular and multimodal datasets \cite{Nam2025}. Semi-autonomous frameworks automate specific stages such as error detection \cite{NarayanCOR22}, entity matching \cite{NarayanCOR22}, or feature engineering \cite{Hollmann0H23}, through iterative code generation and validation. Recent work on semantic operators \cite{lotus25, ovcharenko2026sempipes} further generalizes this paradigm by dynamically delegating fine-grained subtasks to LLMs. Finally, modern frameworks such as skrub unify multiple pipeline variants with different algorithms under a declarative abstraction, exhibiting similar exploratory behavior even without explicit use of agents.

\header{Challenges of \Workload{}:} MLE agents generate large numbers of heterogeneous Python code---from metadata discovery to full pipeline executions---often at a rate of thousands per second \cite{liu26, nvidia2026aiinference}. These agents explore large search spaces by combining broad \emph{exploration} with targeted \emph{exploitation}, yet lack dedicated runtime support, leading to overlapping executions, out-of-memory failures, and inefficient hardware utilization. While prior work proposed optimizations for Text2SQL workloads \cite{liu26, russo26, abs-2508-04031}, these techniques do not transfer to ML workloads. LLM-based query synthesis frameworks primarily improve SQL generation through prompt design \cite{TaiCZ0023, abs-2305-11853}, decomposition \cite{WangR0LBCYZYSL25}, and execution error signals~\cite{ZhaiXHY25}, and rely on mature database engines for efficient execution. In contrast, MLE agents perform guided search over pipeline structures and hyperparameters, repeatedly executing entire pipelines (including model training and evaluation) in a fragmented and inefficient ML ecosystem. As LLM throughput continues to increase, smaller models become more efficient \cite{bitnet, ShahLP25}, specialized hardware for inference advances \cite{freund2026taalas, BaumstarkS26}, and models are explicitly trained for agentic behavior \cite{Hu0XSLLLR25}, research and adoption of autonomy in ML code generation are likely to accelerate. We argue that this shift demands a new class of data systems explicitly designed for \workload{} tailored to ML workloads, for the following reasons:

\header{Insufficient ML Libraries:} MLE agents largely rely on popular Python ML libraries. Over the past decades the Python ML ecosystem has matured significantly, with data scientists heavily depending on high-level libraries such as Pandas and scikit-learn, often combined with a long tail of custom and domain-specific libraries to construct complex pipelines \cite{Psallidasetal2022}. These libraries are deeply embedded in modern ML practice and are extensively represented in LLM pre-training corpora, making them effectively \emph{here to stay} as the primary interface for agent-generated ML code.
However, most ML libraries lack a unifying execution layer that can optimize computation holistically. Instead, they prioritize usability and flexibility over computational efficiency. Moreover, Python itself is not designed for high-performance high-concurrency workloads. As a result, today's Python-based ML ecosystem remains inadequate for supporting the scale of emerging agentic ML workloads.

\header{Lack of Efficient Systems for End-to-end Data Science:} In contrast to mainstream Python libraries, the systems research community made significant advances toward efficient ML execution. First, systems such as SystemML~\cite{BoehmDEEMPRRSST16}, SystemDS~\cite{BoehmADGIKLPR20}, OptiML~\cite{SujeethLBRCWAOO11}, KeystoneML~\cite{SparksVKFR17}, and DAPHNE~\cite{DammeBB0BCDDEFG22} abandon Python in favor of domain-specific languages (DSLs) to enable full-program compilation, cost-based optimization, and multi-backend runtimes, but applying these techniques to the general Python ML ecosystem remains challenging and requires substantial re-engineering. Weld~\cite{palkar2017weld} eases integration but requires manual operator porting and glue code. Despite their technical influence~\cite{Baunsgaard023, PhaniR021}, these systems see limited adoption~\cite{pypistats_systemds, pypistats_daphne}, making them difficult for LLMs to target due to their scarcity in training corpora. 
Second, modern task-specific high-performance systems retain Python interfaces while leveraging optimized native engines. Examples include systems for dataframe processing~\cite{PetersohnMLMXMG20, LuHQLWYLZCD24, polars_2026}, tree-based learning~\cite{ChenG16, KeMFWCMYL17, ProkhorenkovaGV18}, in-process databases \cite{RaasveldtM19}, and DNN workloads~\cite{PaszkeGMLBCKLGA19, AbadiBCCDDDGIIK16}. PyTorch and TensorFlow also support JIT compilation~\cite{TorchDynamo, MoldovanDWJLNSR19},  though these features remain underutilized~\cite{alawi2025comparativesurveypytorchvs, you2025depyf}. Distributed frameworks like Ray~\cite{MoritzNWTLLEYPJ18} and Dask~\cite{Dask-scipy-2015} focus on task scheduling rather than holistic optimization, while AutoML frameworks~\cite{erickson2020autogluontabularrobustaccurateautoml, LibertyKXRCNDSA20, abs-2007-04074} automate model selection but lack compilation and runtime support. However, these optimizations are largely specialized to specific tasks and do not generalize across heterogeneous pipelines without additional research effort.
Despite these advances, the ecosystem lacks a unified system that efficiently executes heterogeneous, end-to-end data science pipelines within Python and provides a common abstraction for optimization across pipeline stages. The current landscape remains fragmented---optimized for isolated tasks---leaving a substantial gap between the usability of Python APIs and the performance demands of large-scale, agentic ML workloads.

\begin{figure}[!t]
	\centering
        \includegraphics[scale=1.1]{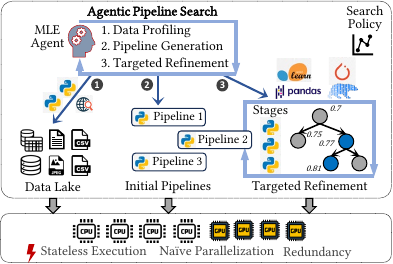}
	\caption{\label{fig:agent} \Workload{}} 
	\vspace{-0.1cm}
\end{figure}

\header{Our Vision:} We envision a new ML system designed to efficiently execute large-scale  \workload, addressing the fundamental mismatch between high-level Python APIs and low-level optimizations. To this end, we introduce \emph{stratum}, a system that integrates with MLE agents to accelerate \workload{}. Stratum\footnote{stratum:\textcolor{blue}{https://github.com/deem-data/stratum}} raises the scope of optimization from individual pipelines to the entire search process. It represents batches of agent- or user-generated pipelines---written using popular or custom ML libraries---as lazily evaluated directed acyclic graphs (DAGs), without requiring DSLs or manual operator porting, applies logical rewrites, lowers logical operators to physical implementations, and executes them across heterogeneous backends, including a novel Rust-based runtime. Our detailed contributions are:
\begin{itemize}[leftmargin=*]
    \item \emph{Agentic Workloads:} We present a representative use case for \workload{} and analyze the execution characteristics of agentic workloads, motivating the design of stratum (\Cref{sec:usecase}).
    \item \emph{Design Principles:} We discuss our design principles for supporting large-scale \workload{} (\Cref{sec:vision}).
    \item \emph{System Architecture:} We describe stratum's architectural vision, including its logical optimizer, operator selection, Rust backend, parallelization planning, and reuse of intermediates (\Cref{sec:sysarch}). Stratum integrates seamlessly with arbitrary ML libraries.
    \item \emph{Challenges and Key Directions:} We discuss the status of our \emph{early prototype}, open challenges and research directions~(\Cref{sec:directions}).
    \item \emph{Preliminary Experiments:} Using a real-world workload generated by the AIDE agent~\cite{abs-2502-13138}, we present preliminary results from both end-to-end and microbenchmark evaluations~(\Cref{sec:exp}). Even in its early implementation, stratum yields significant speedups over the baselines, validating our design decisions.
\end{itemize}


\section{Agentic ML Workloads} \label{sec:usecase}

We introduce a representative enterprise use case for agentic ML to discuss scale and characteristics of this type of workloads.

\header{End-to-end \Workload{} Use Case:} Consider an enterprise data scientist building a customer churn model from heterogeneous datasets stored in a shared data lake, including CSV and Parquet files (demographics, transaction and usage logs) as well as text and images accumulated over years. The data scientist invokes an MLE agent to generate an end-to-end executable pipeline.
As shown in Figure~\ref{fig:agent}, the agent first profiles the data---extracting data characteristics, missing value ratios, cardinalities, vocabulary diversity, and image metadata---often via repeated sampling~\cite{abs-2509-21825, fang2025mlzero}. Based on the inferred problem type and data characteristics, it retrieves relevant approaches from the web~\cite{Nam2025} and synthesizes candidate pipelines combining alternative preprocessing strategies, model families, text-derived features, and multimodal variants using specialized libraries. The agent executes and evaluates these pipelines, prunes infeasible candidates, and then shifts from broad \emph{exploration} to targeted refinement (\emph{exploitation}) by iteratively modifying pipeline stages (e.g., feature transformation, imputation, model training), exploring stage-wise variants and hyperparameters~\cite{Nam2025}. This refinement–evaluation loop repeats across pipeline variants, followed by constructing ensembles, and cross-validation. The data scientist may rerun the agent with modified prompts or launch multiple agents in parallel, each executing pipelines in separate Python processes, to broaden the search.

\begin{figure}[!t]
    \centering
    \subfigure[Changed Code Lines]{
        \label{fig:code-changes}
        \includegraphics[width=0.39\columnwidth]{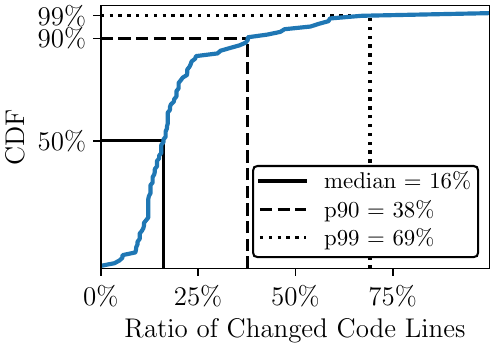}}
    \hfill
    \subfigure[System Utilization]{
        \label{fig:system-utilization}
        \includegraphics[width=0.56\columnwidth]{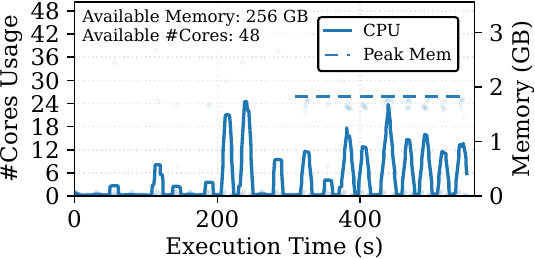}}
    \vspace{-0.20cm}
    \caption{\label{fig:agentic-search}Distribution of code changes and CPU/Memory utilization during agentic search for an example workload.}
    \vspace{-0.1cm}
\end{figure}

\header{Execution Characteristics of Agentic ML Workloads:} This use case exposes several execution patterns typical of agentic ML: 
(1) \emph{Redundant Computation:} Dataset discovery triggers many lightweight profiling scripts; for large datasets, repeated data loading often dominates cost. During targeted refinement (exploitation phase), agents execute large numbers of pipelines repeatedly applying similar transformations, resulting in significant redundant computation. Figure~\ref{fig:code-changes} highlights this on a Kaggle competition~\cite{uk_housing_prices_paid} executed by the AIDE~\cite{abs-2502-13138} agent: 50\% of iterations modify 16\% or fewer lines of code, indicating substantial overlap across pipeline variants.
(2) \emph{Limited and Process-based Parallelism:} Modern agents \cite{Nam2025, abs-2502-13138} execute Python code in a largely stateless manner, maintaining exploration state (e.g., accuracy feedback and search strategy) outside the runtime. Pipelines are therefore evaluated sequentially or via naive parallelization, spawning many independent Python processes. This incurs substantial hardware contention and serialization overhead. Coarse-grained scale-out further increases cost and energy consumption~\cite{abs-2507-02554}.
(3) \emph{System-agnostic Execution:} Agents operate without awareness of underlying system characteristics such as available memory, CPU cores, GPUs, or distributed backends, leading to out-of-memory (OOM) failures, underutilized hardware, and missed opportunities for holistic data- and task-parallel execution and intermediate reuse. Figure~\ref{fig:system-utilization} shows highly irregular CPU and memory utilization, indicating poor resource efficiency during agentic search.
These inefficiencies are most visible in fully autonomous agents but also arise in semi-autonomous settings. Together, they result in poor resource utilization, redundant materialization and computation, and overall inefficient execution.


\section{Our Vision} \label{sec:vision}
We motivate the necessity for a new system tailored to \workload{} and outline its key technical requirements.

\header{The Case for a New System:}
As LLMs are increasingly used for code generation, the popularity of mainstream and specialized Python libraries will continue to grow. Modern MLE agents employ diverse search policies~\cite{abs-2506-13131, fang2025mlzero} to navigate large search spaces of independently evolving libraries and APIs. These include running multiple agent instances with different prompts~\cite{abs-2506-13131}, exploring sampled datasets~\cite{abs-2505-21372}, naive parallelization across nodes and processes~\cite{Nam2025, abs-2506-13131}, and restarting exploration based on early feedback~\cite{abs-2505-21372}. However, they primarily optimize search strategies to maximize pipeline quality (e.g., accuracy), while largely overlooking execution latency and throughput. As discussed in Section~\ref{sec:intro}, existing ML systems either target narrow workload classes or rely on DSLs that are ill-suited for LLMs. As a result, the scale of \workload{} cannot be met through incremental extensions to existing systems. We argue that the sheer diversity of ML libraries, coupled with the absence of common abstractions necessitates a new class of systems designed for agentic ML workloads.

\header{The Vision for Stratum:}
To this end, we envision \emph{stratum}, a system architecture for large-scale \workload{}. 
Specifically, our design is based on the following key principles:
\begin{itemize}[leftmargin=*]
    \item \emph{Seamless and unrestricted support} for ML libraries (scikit-learn, Pandas), frameworks (PyTorch, TensorFlow), tabular foundation models~\cite{Hollmann0EH23, QuHVM25}, and specialized libraries---without requiring operator porting---while remaining extensible to future libraries.
    \item \emph{A semantic abstraction} built on a minimal set of logical operators that enables rewrites (e.g., common subexpression elimination (CSE)) and lazy evaluation with physical operator independence.
    \item \emph{A runtime} with efficient operator kernels and scheduling across heterogeneous backends including CPUs, GPUs, and distributed backends to support diverse data characteristics and ML tasks, and cost-based runtime optimizations including reuse of intermediates, memory management, and parallelization planning.
\end{itemize}
However, realizing this vision raises a range of open challenges across representation, optimization, and execution, and demands novel techniques throughout the system stack.

\begin{figure}[!t]
	\centering
        \includegraphics[scale=0.68]{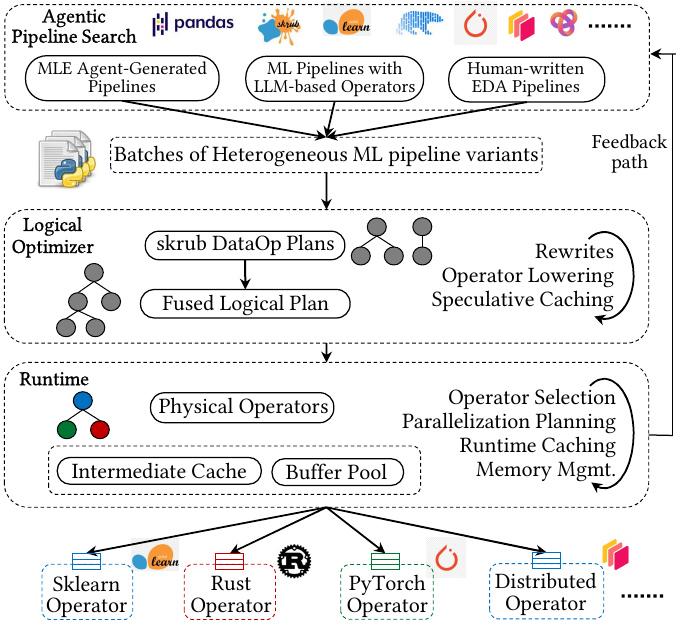}
	\caption{\label{fig:arch}Stratum System Architecture}
	\vspace{-0.1cm}
\end{figure}

\section{System Architecture}\label{sec:sysarch}

Here we describe the overall architecture of stratum (Figure~\ref{fig:arch}) and its key components. Stratum builds on skrub's \cite{skrub} operator abstractions. At a high level, stratum ingests batches of ML pipelines, fuses them into a unified operator DAG, applies logical optimizations and operator lowering, and selects efficient backends for execution. We implement a Rust-based execution engine along with system-level optimizations such as nested parallelism and intermediate caching. While many of these techniques draw inspiration from prior research, our primary contribution lies in the principled integration of these techniques into a holistic system underneath the Python ML ecosystem, which we argue is both timely and of utmost necessity.

\subsection{A Declarative Abstraction} \label{sec:skrubabs}
Logical and runtime optimizations require representing ML pipelines as a DAG with well-defined operator semantics, which is challenging due to the lack of a universal abstraction for heterogeneous ML code~\cite{GrafbergerGSS22}. Translating arbitrary Python ML code into a 
\setlength{\intextsep}{5pt}
\setlength{\columnsep}{6pt}
\begin{wrapfigure}{r}{5.1cm}
	\centering
        \includegraphics[scale=0.65]{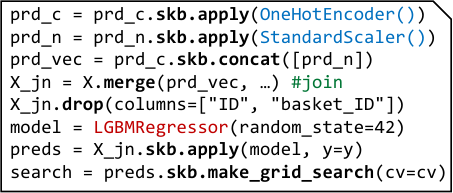}
	\vspace{-0.1cm}
	\caption{\label{fig:skrub_code}Example Skrub Code}
\end{wrapfigure}
DAG via instrumentation is also difficult given Python's dynamic semantics \cite{GrafbergerGGS23, LiCSPS24, FangCSP25}. 
Stratum adopts \emph{skrub}~\cite{skrub} as its entry point because it converts arbitrary ML pipelines into lazily evaluated DAGs. Skrub automatically wraps operations---including Pandas, scikit-learn, DNN models, custom library calls, and UDFs---into semantically explicit operators (\texttt{DataOps} \cite{dataops_API}), forming a control-flow-free execution DAG. This abstraction unifies data preparation, model training, and cross-validation constructs into a single lazily executable pipeline. Figure \ref{fig:skrub_code} shows a simplified skrub pipeline, where \texttt{make\_grid\_search} forms and executes the DAG with cross-validation. Although designed for usability rather than performance, skrub's operator semantics provides a foundation for system-level optimization. 
Stratum extends these abstractions with new operators, iterative lowering, and its own compiler and runtime stack, while remaining API-compatible with skrub.

\subsection{Logical Optimizer and Runtime} \label{sec:optimizer}
Agents emit pipeline variants in batches. Stratum fuses each batch into a unified DAG and passes it to the logical optimizer.

\header{Logical Operators and Rewrites}: We compile skrub DataOps into a common logical operator hierarchy (e.g., SELECT, PROJECT, MAP, TRANSFORMER, and ESTIMATOR). Skrub treats most operators as black boxes. Stratum performs a metadata collection pass to extract operator-level metadata, such as operator type (e.g., dataframe operation, estimator), source library, structural properties (e.g., selection and projection), and data characteristics (e.g., \#rows, \#cols, and datatypes). Using this metadata, the optimizer applies logical rewrites that preserve semantic equivalence, including predicate pushdown, read sharing, CSE, and constant folding  
\setlength{\intextsep}{1.1pt}
\setlength{\columnsep}{6pt}
\begin{wrapfigure}{r}{4.5cm}
	\centering
        \includegraphics[scale=0.58]{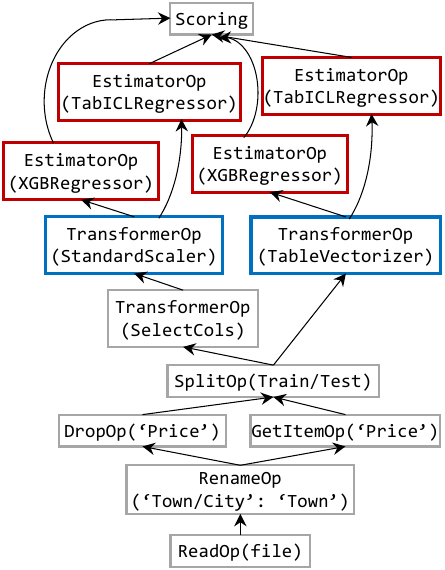}
	\vspace{-0.1cm}
	\caption{\label{fig:dag}Example DAG}
    \vspace{-0.05cm}
\end{wrapfigure}
%
to reduce redundant computation and improve data locality. Figure~\ref{fig:dag} shows a simplified execution graph after fusing four pipelines that combine two preprocessing techniques (in {\color{blue}blue}) with two models (in {\color{red}red}) and applying logical rewrites. Stratum also applies API-aware rewrites, such as reordering Pandas operations to reduce copies under copy-on-write, enabling in-place updates, and replacing non-vectorized loops with vectorized implementations.

\header{Operator Lowering:} Following metadata collection and logical rewrites, stratum lowers logical operators (e.g., TRANSFORMER, ESTIMATOR) into fine-grained physical operators (e.g., \texttt{OneHotEncoder} or \texttt{ElasticNet}) and apply physical-level rewrites. It further decomposes high-level operators such as skrub's \texttt{TableVectorizer} (Figure~\ref{fig:dag}), which encapsulates automatic missing value imputation and feature transformations, into independent operators like \texttt{cleaner}, \texttt{StringEncoder}, and \texttt{OneHotEncoder}. Similarly, higher-level constructs such as cross-validation and hyperparameter search, which repeatedly execute the same subgraph, are unrolled into explicit DAGs. This lowering enables accurate cost estimation, runtime planning, and memory management.

\header{Operator Selection:} Stratum maintains multiple implementations for each physical operator with distinct algorithmic and computational characteristics. For example, dimensionality reduction may execute via scikit-learn's SVD, an approximate alternative such as Frequent Directions~\cite{Huang19} during early exploration, or stratum's native Rust runtime. We maintain a registry of available implementations and select them via late binding. This logical--physical separation enables extensibility across new libraries (e.g., PyTorch, Spark), execution backends (e.g., Dask for out-of-core execution), custom operators, and hardware environments, while naturally supporting both lazily and eagerly evaluated frameworks.
Using collected metadata and compute and memory estimates, stratum selects operator implementations that minimize execution time under memory constraints. The optimal choice depends on neighboring operators, data placement, and the availability of intermediates in the cache. For example, if downstream PyTorch operators are placed on GPUs, stratum may prefer cuDF over Pandas or Polars to avoid data movement. For operators that fit in-memory, we prefer efficient native (GIL-releasing) backends such as Rust, Polars, cuDF, or NumPy-backed implementations.

\header{Rust Backend:} While efficient frameworks exist for specific ML stages (e.g., Polars, XGBoost, PyTorch), many widely used operators in scikit-learn, and Pandas are implemented in Python. Even when backed by NumPy or Cython, their highly generic implementations---designed to support many hyperparameters and edge cases---incur repeated type conversions, temporary allocations, unnecessary copies, and limited multithreading. Such overheads often dominate end-to-end execution time, diminishing the gains from optimizations (Amdahl's Law).
To address this, we incrementally develop Rust implementations for frequently used operator configurations~\cite{Psallidasetal2022}, exposed via lightweight \texttt{PyO3} bindings \cite{johnson2025pyo3}. Rust enables efficient kernels with explicit memory control and zero-copy data access, while fitted model states remain in Rust-managed objects to eliminate data conversions. These kernels release Python's global interpreter lock (GIL) to enable concurrent kernel execution and exploit native data-parallelism. In addition, the Rust backend lowers operator granularity, enables accurate cost estimates, improves memory management, reduces boundary crossings, and enables advanced optimizations such as operator fusion~\cite{BoehmRHSEP18}, sparsity exploitation~\cite{Sommer0ERH19}, and fine-grained reuse \cite{PhaniR021}.

\subsection{Runtime Optimizations} \label{sec:runtime}

\header{Inter- and Intra-operator Parallelism:} A major performance bottleneck in ML libraries stems from Python’s limited support for native parallelism. While multithreading is standard in systems, Python’s GIL restricts concurrency unless native kernels explicitly release it. Multiprocessing-based schedulers (e.g., Dask, Ray, Joblib) achieve parallelism for scikit-learn estimators, but incur high memory and serialization overhead due to process-level data duplication. Moreover, many ML libraries---and stratum’s Rust backend---already employ internal multithreading (e.g., OpenMP, OpenBLAS, Rayon), introducing nested parallelism that risks oversubscription. Although optional GIL removal exists in recent Python versions~\cite{python_freethreading_2026}, adoption remains limited. 
Stratum primarily relies on multithreading. During the exploration phase, agents evaluate many non-overlapping pipelines, increasing opportunities for inter-operator parallelism. Using compute and memory estimates, and available hardware, a cost-based optimizer traverses the DAG, evaluates plans under worst-case memory budgets, and selects plans that minimize execution time subject to memory constraints~\cite{ZhengLZZCHWXZXG22, PhaniEB22}. In a subsequent pass, we determine the degree of intra- and inter-operator parallelism to avoid oversubscription.

\header{Reuse of Intermediates:} To exploit the repetitive nature of the exploitation phase of \workload{} (see Figure~\ref{fig:code-changes}), we employ a combination of coarse-grained reuse \cite{XinMMLSP18}---caching results of top-level operators---with fine-grained reuse \cite{PhaniR021, Phani025, PhaniThesis25, PhaniB26} across shared Rust kernels. The cache maps operator hashes to materialized outputs (e.g., dataframes, NumPy arrays, Rust-backed model states). Operator hashes are derived from input hashes and operator specifications; seed values (e.g., \texttt{random\_state}) are incorporated to handle non-determinism, while operators without explicit seeds are not cached. Before execution, each operator probes the cache and reuses previously materialized results when available. 
We allocate a fixed fraction of memory (default 10\%) for cached objects. To assist the runtime caching, the optimizer speculatively marks selected operators (e.g., expensive preprocessing) as cache candidates. Cached intermediates are materialized to Parquet and lazily reloaded across iterations. Unlike prior reuse frameworks, designed for iterative human-driven ML workflows or DSL-based pipelines~\cite{XinMMLSP18, PhaniR021, Phani025}, agentic workloads exhibit far larger and more dynamic exploration spaces spanning profiling, heterogeneous pipelines, and DNN workloads. Moreover, exploration-exploitation patterns and multi-tenant execution (see Section~\ref{sec:directions}) make reuse decisions substantially more challenging \cite{Baunsgaard0IKLO22}, motivating specialized caching and reuse strategies for agentic ML workloads.


\section{Challenges and Key Directions}
\label{sec:directions}
This section first summarizes the current state of the prototype and then discusses major challenges and long-term directions.

\header{Prototype Status:} The current prototype establishes stratum's core architecture, including an initial logical optimizer with metadata collection and basic rewrites. The Rust backend provides a limited set of execution kernels sufficient to validate key design choices. At present, stratum supports only in-memory operators and relies on heuristics for operator selection and parallelization planning.  Finally, we employ a greedy caching strategy that materializes expensive preprocessing operators for reuse. These initial implementations serve as placeholders for more principled cost-based and adaptive strategies, which are under active development.

\header{Open Challenges:} While our early prototype shows promising results, several open challenges remain. (1) \emph{Cost Estimates:} Operator selection and runtime optimizations require decent memory and compute estimates, which are difficult to obtain as Pandas and scikit-learn methods often materialize hidden intermediates that inflate  memory usage. We are exploring sampling-based cost estimation~\cite{DuboutF11} and adaptive planning.
(2) \emph{Cross-library Optimizations:} Heterogeneous pipelines combine operators from diverse libraries (e.g., Pandas, NumPy, PyTorch) with widely different execution granularities, from lightweight dataframe transformations to expensive trainings. These boundaries restrict cross-library fusion and complicate scheduling and inter-operator parallelism. We address this by increasingly relying on homogeneous Rust-based operators, which enable more effective fusion, scheduling, and parallelization.
(3) \emph{UDFs:} Custom pipelines components and specialized libraries are often wrapped as UDFs, creating black-box operators that hide semantic information. We are exploring profiling on sampled inputs to estimate costs and adaptive optimizations based on observed costs. For common map-style functions, output dimensions are often preserved, enabling accurate memory estimation.
(4) \emph{DNN Workloads:} Supporting DNN and agentic kernel generation requires deeper CUDA and PyTorch integration~\cite{Ouyang0AZHRM25, xu2026vibetensorsoftwaredeeplearning}. We are exploring skorch~\cite{skorch} as well as Rust bindings to PyTorch (e.g., \texttt{tch-rs}) to enable fine-grained optimizations such as quantization, weight pruning, input data pipeline reuse, and pipeline parallelism.

\header{Agent-System Co-design:} 
Stratum integrates with existing agents; however, a tighter integration between agents and stratum exposes opportunities for jointly optimizing the agentic search and its execution:
(1) \emph{Workload-aware Optimizations:} Instead of executing independent Python scripts, agents annotate pipelines with lightweight metadata (e.g., exploration vs. exploitation stage), enabling stratum to adapt execution strategies such as successive halving~\cite{LiJDRT17}, reduced training epochs, lower-fidelity operators during exploration, and compiler-assisted caching of operators with high reuse potential during exploitation. 
(2) \emph{Declarative Pipeline Specification:} Instead of generating full Python scripts, system-aware agents emit abstract pipeline specifications~\cite{abs-2506-13131}, allowing stratum to internally construct execution graphs. This reduces the number of LLM calls and energy consumption, while expanding optimization opportunities. 
(3) \emph{Overlapping Generation and Execution:} Agents generate pipelines in batches and wait for feedback. Stratum mitigates idle time by executing subsets of each batch, returning early feedback, and overlapping execution with subsequent pipeline generation. 
Together, these mechanisms elevate optimization from individual pipeline executions to the entire search process, motivating a new class of agents that jointly optimize pipeline search and execution.

\header{Other Key Directions:} Beyond building a robust system implementation and a representative system-aware agent, several longer-term research directions remain: (1) \emph{Multi-tenancy:} We envision stratum evolving into a multi-tenant, cloud-hosted service that interfaces directly with MLE agents, enabling dynamic scheduling, adaptive resource management, and shared intermediate caches across workloads. (2) \emph{Inference engine:} A complementary direction is tailoring the inference stack for agentic workloads through customized inference configurations and shared key-value caches, further reducing latency and cost across agent interactions~\cite{ZhengYXS0YCKSGB24}.


\section{Preliminary Experiments}
\label{sec:exp}

We evaluate the baseline performance of stratum's early prototype and its core components to validate its key design principles.

\header{HW Environment and Workload:} We run all experiments on a single node with an AMD EPYC 7443P CPU (24 / 48  cores) and 256$\gb$ RAM, using Ubuntu~20.04, Python~3.11, scikit-learn~1.8, and skrub~0.6.2.
We use AIDE~\cite{abs-2502-13138} to generate a representative \workload{} workload with two iterations. The first iteration explores all combinations of two preprocessing strategies and four models. The preprocessing strategies include: (1) missing-value imputation and feature encoding using \texttt{StringEncoder}, a custom target encoder, and \texttt{StandardScaler}; and (2) \texttt{TableVectorizer}, which performs automatic cleaning and applies one-hot encoding and \texttt{StringEncoder} to low- and high-cardinality features. The models include \texttt{Ridge}, \texttt{XGBoost}, \texttt{LightGBM}, and \texttt{ElasticNet}. In the second iteration, we select the best-performing preprocessing strategy and model based on validation accuracy and perform hyperparameter tuning. We use the UK housing dataset from Kaggle~\cite{uk_housing_prices_paid} and vary dataset sizes to evaluate scalability.

\begin{figure}[!t]
	\centering
	\subfigure[End-to-end Performance]{
		\label{fig:end2end}\includegraphics[scale=0.39]{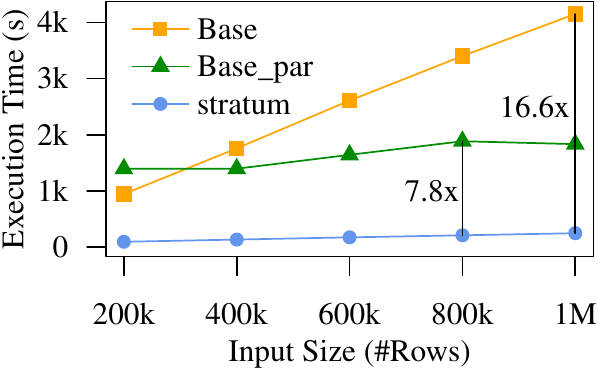}}
	\subfigure[Optimization Breakdown]{
		\label{fig:micro}\includegraphics[scale=0.39]{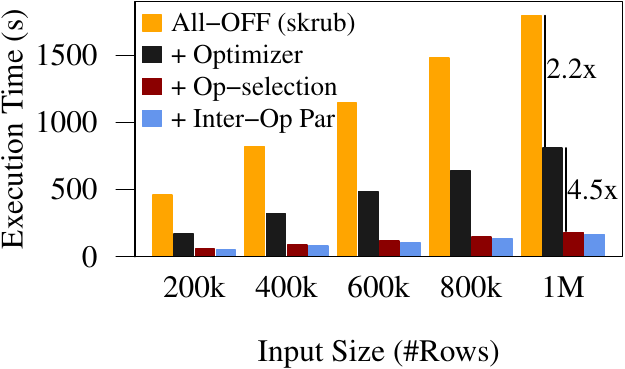}}
    \vspace{-0.25cm}
	\caption{\label{fig:exp}Impact of stratum and its optimizations.}
    \vspace{-0.1cm}
\end{figure}

\header{End-to-end Performance:} We compare AIDE (which performed best in MLEBench) with stratum. The baselines include: \textbf{Base}, representing AIDE with sequential pipeline execution; \textbf{Base\_par}, where AIDE triggers multiple pipelines concurrently; and stratum with all optimizations enabled. 
As shown in Figure~\ref{fig:end2end}, stratum yields a  16.6$\times$ speedup over Base. This improvement stems from: (i) pipeline fusion in the first iteration, (ii)~CSE to deduplicate preprocessing, (iii)~operator selection (Polars over Pandas and our Rust kernels over scikit-learn), (iv)~intra- and inter-operator parallelism, and (v)~reuse of preprocessing results in the second iteration. While Base\_par improves upon Base, its reliance on multiprocessing incurs significant serialization overhead and increases memory consumption by 8$\times$. Stratum remains 7.8$\times$ faster than Base\_par.

\header{Ablation Study:} To study the impact of stratum’s individual optimizations---which vary with data and workload characteristics---we incrementally enable each optimization in isolation. As shown in Figure~\ref{fig:micro}, logical optimization alone yields up to a 2.2$\times$ speedup through CSE and related rewrites. Enabling operator selection provides an additional 4.5$\times$ improvement by replacing Python-based operators with native implementations, which also enable data-parallelism by releasing the GIL and leveraging multi-threading. Finally, inter-operator parallelism contributes a further 10\% speedup. In this workload, the dominant operators are already compute-intensive and fully utilize all available cores, limiting the additional gains achievable through inter-operator parallelism.

Overall, these results validate our design principles, showing that even an early prototype of stratum can significantly accelerate workloads through holistic logical and runtime optimizations.

\section{Additional Related Work}
Beyond the prior work discussed in Section~\ref{sec:intro}, stratum is related to ML systems, accelerating dataframes, and agentic SQL workloads.

\header{Scalable Data Science:} Systems for accelerating data science~\cite{PetersohnMLMXMG20, LuHQLWYLZCD24, EmaniFC24, HagedornKS21, SinghKB026, JindalEDPHPG0CM21, SchuleSK023, SinghKB026} improve dataframe performance by rule-based and dynamic tiling, parallel execution, or by translating Pandas operations to SQL. Our logical rewrites draw inspiration from mlwhatif~\cite{GrafbergerGGS23}, which builds an operator DAG by instrumenting Python code. In contrast, stratum constructs a lazily evaluated operator DAG directly from arbitrary ML libraries and applies advanced optimizations including rewrites and operator selection.

\header{ML Systems Optimizations:} Our compiler and runtime techniques relate to prior work on ML systems~\cite{BoehmADGIKLPR20, BoehmDEEMPRRSST16, KunftKSBRM19}, pipeline parallelism~\cite{HuangCBFCCLNLWC19, FanRMCWZWLYXDLL21}, inter- and intra-operator parallelism~\cite{ZhengLZZCHWXZXG22}, task-based execution~\cite{PhaniEB22, BoehmTRSTBV14, MoritzNWTLLEYPJ18}, and coarse- and fine-grained reuse~\cite{XinMMLSP18, PhaniR021, Phani025, PhaniThesis25}. In contrast, stratum employs operator scheduling, multi-level parallelism, and reuse for \workload{}.

\header{Agentic SQL:} Stratum is also related to systems for Text2SQL \cite{liu26, LiuBKCSPCMSGZ25, abs-2508-04031} and optimizing semantic operators~\cite{russo26, sun2025agenticdataagenticdataanalytics, lotus25, LiuRC0CCFK0SV25, JoT24, DorbaniYLM25}. Unlike these systems, stratum targets agentic ML workloads.




\section{Conclusion}

By elevating ML development from manual scripting to autonomous generation and iterative refinement, \workload{} introduces a fundamentally new workload pattern that necessitates a new class of systems tailored to agentic ML workloads. This raising of abstraction in ML development---together with the ML community’s growing appreciation for high-level operator semantics~\cite{skrub}---creates an opportunity to rethink end-to-end ML system design. We introduced stratum, a system infrastructure for \workload{} that enables logical and runtime optimizations while remaining fully compatible with existing Python libraries. In summary, stratum comprises (1)~an execution engine supporting lazy evaluation, (2)~an efficient Rust backend, (3)~cost-based operator selection, parallelization, and caching, and (4)~an optimizer that abstracts heterogeneous ML libraries under a unified execution model.


\bibliographystyle{ACM-Reference-Format}
\bibliography{vldb}

\end{document}